\documentclass[preprint, pre]{revtex4}
\usepackage{graphicx}
\usepackage{subfigure}
\usepackage{amsmath,amsfonts}
\DeclareGraphicsExtensions{.eps}
\graphicspath{{c://users/srkim_000/documents/matlab/}}

\begin{document}
%%%%%%%%%%%%%%%%%%%%%%%%%%%%%%%%%%%%%%%%%%%%%%%%%%%%%%%%%%%%%%%%%%%%%%%%%%%%
\title{A New Description of Quantum Behaviors for a Simple Harmonic Oscillator}
%%%%%%%%%%%%%%%%%%%%%%%%%%%%%%%%%%%%%%%%%%%%%%%%%%%%%%%%%%%%%%%%%%%%%%%%%%%%

\author{Sangrak Kim}
\address{Department of Physics,\\ Kyonggi University,\\ 154-42 Gwangyosanro, Youngtong-ku, Suwon 440-760, Korea}
\date{\today}
%
%%%%%%%%%%%%%%%%
\begin{abstract}
%%%%%%%%%%%%%%%%
We describe quantum behaviors of a simple harmonic oscillator, starting from the classical mechanics. By imposing two conditions on the phase points generated from a symplectic algorithm, we obtain discrete energy levels, satisfying $E_n \tau_n = h$ where $E_n$ is the total energy of the oscillator and $\tau_n$ is the time step for the closed orbit of $n$-polygon in phase space. We can thus successfully integrate classical and quantum mechanics into a single frame, if we assume that time is discrete.

%%%%%%%%%%%%%%%%
\end{abstract}
%%%%%%%%%%%%%%%


\maketitle

A simple harmonic oscillator is one of the simplest and best-known models in physics. It can be exactly solved not only classically but also quantum mechanically, so it appears in almost all the general physics textbooks \cite{text} at the undergraduate level. Nevertheless, we can further find out a plenty of interesting facts about it and thus, it is a problem still to be solved. The Hamiltonian of the simple harmonic oscillator is written as
\begin{equation}
 H(q, p) = \frac{1}{2} p^2 + \frac{1}{2}kq^2 \label{tripot}
\end{equation}
where $k$ is the spring constant, $q$ is the coordinate of the particle from its equilibrium point, and $p$ is its momentum. Here, we assume that the particle has unit mass, for simplicity. Classically, the particle moves on an ellipse in phase space with $H=p^2/2+ kq^2/2= \textrm{constant}$. Quantum mechanically, it has discrete energy levels that are given by $E_n = (n+1/2) \hbar \omega$ where $\omega =\sqrt{k}$ is the angular frequency of the oscillator and $\hbar = h/2\pi$ is the reduced Planck constant. Note that the energy levels are evenly distributed as $\Delta_n \equiv (E_{n+1} -E_n)/\hbar \omega = 1$.

In this article, we will examine the possibility of describing the quantum behaviors of the simple harmonic oscillator, starting from the framework of classical molecular dynamics. If we apply, for example, a symplectic Euler algorithm to the Hamiltonian $H$ in Eq. (\ref{tripot}), the evolution of the phase points can be simply written as

\begin{equation}
\left[ \begin{array}{c} q_{n+1}\\ p_{n+1} \end{array} \right] = A \left[ \begin{array}{c} q_n \\ p_n \end{array} \right],
\label{eulereq1}
\end{equation}
where the transformation matrix is given as
\begin{equation}
A = \begin{bmatrix} 1 & \tau \\ -k\tau & 1-k\tau^2 \end{bmatrix},
\label{symat2}
\end{equation}
and $\tau$ is an arbitrary time step. A modified equation for the Hamiltonian $H$ \cite{Toxvaerd} represents the discrete form of the continuous Hamiltonian equations and the discrete phase trajectory moves on the surface of the modified Hamiltonian $\tilde{H} = \textrm{constant}$. From Eqs. (\ref{tripot})-(\ref{symat2}), we obtain the corresponding modified Hamiltonian as
\begin{equation}
    \tilde{H} (p, q)= f(\theta)(p^2+kq^2+k\tau pq),
\notag
\label{modifiedHamiltonian2}
\end{equation}
where $f(\theta) \equiv 2\sin^{-1}{\sqrt{\theta/2}}/\sqrt{4\theta-\theta^2}$ and $\theta \equiv k\tau^2$. Note that $\theta$ is a dimensionless parameter with $0 < \theta<4$. We can also rewrite Eq. (\ref{eulereq1}) as

\begin{equation}
\left[ \begin{array}{c} q_{n}\\ p_{n} \end{array} \right] = A^n \left[ \begin{array}{c} q_0 \\ p_0 \end{array} \right].
\notag
\label{geneq1}
\end{equation}
If we set
\[ A^n = \left( \begin{array}{cc}
M_{11} & M_{12} \\
M_{21} & M_{22} \end{array} \right),\]
then we obtain

\begin{subequations} \label{eq:ms}
\begin{align}
M_{11} &= \frac{(1-\frac{\theta}{G}) (1-\frac{G +\theta}{2})^n + (1+\frac{\theta}{G}) (1+\frac{G -\theta}{2})^n}{2}, \notag \\
M_{12} &= - \tau \frac{(1-\frac{G +\theta}{2})^n - (1+\frac{G -\theta}{2})^n}{G},\notag  \\
M_{21} &= k\tau \frac{(1-\frac{G +\theta}{2})^n - (1+\frac{G -\theta}{2})^n}{G}, \notag \\
M_{22} &= \frac{(1+\frac{\theta}{G}) (1-\frac{G +\theta}{2})^n + (1-\frac{\theta}{G}) (1+\frac{G -\theta}{2})^n}{2}, \notag
\end{align}
\end{subequations}
where $G(\theta) \equiv \sqrt{\theta (\theta-4)}$.

Now, let us first impose the stationary condition, $A^n = I$, for each $n$ where $I$ is the identity matrix. Then, the phase points actually make a closed orbit in the phase space with a cycle $n$. In other words, the trajectory repeats itself at every $n$ step. This implies that the initial total energy of the system is conserved in every $n$ step. We can determine the value of $\theta_n$ and thus obtain the discrete time step $\tau_n =\sqrt{\theta_n}/\omega$ for each cycle $n$. Here, we include the subscript $n$, in order to explicitly denote their dependency on the cycle number $n$. The calculated results for $\theta_n$ and $\tau_n$ are summarized in Table \ref{tab:01} for the first several values of $n$. Actually, higher values of $n$ were calculated, but are not shown, since they are rather complex and lengthy to fit in Table \ref{tab:01}.

\begin{table}
\caption{\label{tab:01} Summary of calculated results.}
 \begin{tabular}{||c|c|c|c|c|c|c|c|c|c|c|c||} \hline
 $n$        & $\theta_n$   & $\tau_n \omega$   & $q_n/\sqrt{2h/\omega}$   & $E_n/h \omega$  \\ \hline
 3        & 3 & $\sqrt{3}$ & $1/\sqrt[4]{3}$ & $1/\sqrt{3}$ \\ \hline
 4        & 2 & $\sqrt{2}$ & $1/\sqrt[4]{2}$ & $1/\sqrt{2}$ \\ \hline
 5        & $(5-\sqrt{5})/2$ & $\sqrt{(5-\sqrt{5})/2}$ & $1/\sqrt[4]{(5-\sqrt{5})/2}$ & $1/\sqrt{(5-\sqrt{5})/2}$ \\ \hline
 6        & 1 & 1 & 1 & 1 \\ \hline
 \end{tabular}
\end{table}

\begin{figure}[htbp]
\begin{center}
\includegraphics[scale=.7]{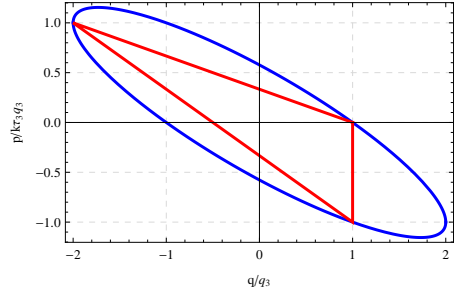}
\caption{(Color on-line) Phase trajectory with cycle $n=3$. Coordinates are shown in reduced units. The closed orbit makes a triangle. The enclosed area of the triangle is $\oint p_ndq_n = \frac{3}{2}k \tau_3 q_3 ^2$. The phase point starts from $(q', p')=(1, 0)$ and rotates clockwise on the ellipse with $\tilde{H}=$ constant.}
\label{fig:01}
\end{center}
\end{figure}

\begin{figure}[htbp]
\begin{center}
\includegraphics[scale=.7]{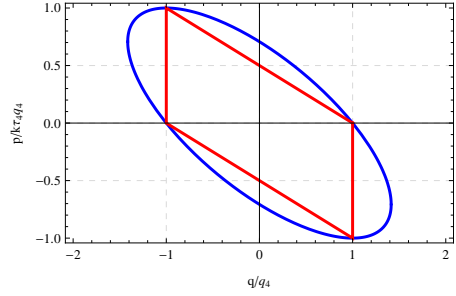}
\caption{(Color on-line) Phase trajectory with cycle $n=4$. The notation is the same as in Fig. \ref{fig:01}. The enclosed area of the tetragon is $\oint p_ndq_n = \frac{4}{2}k \tau_4 q_4 ^2$.}
\label{fig:02}
\end{center}
\end{figure}

\begin{figure}[htbp]
\begin{center}
\includegraphics[scale=.7]{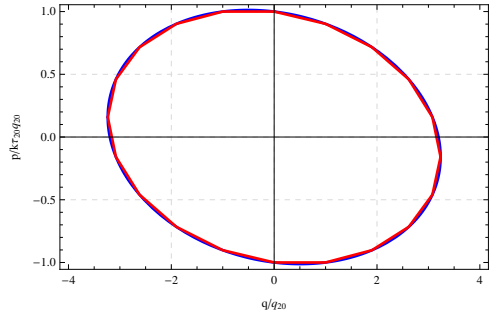}
\caption{(Color on-line) Phase trajectory with cycle $n=20$. The notation is the same as in Fig. \ref{fig:01}. The enclosed area of the polygon is $\oint p_ndq_n = \frac{20}{2}k \tau_{20} q_{20} ^2$. As $n$ becomes larger, the $n$-polygon becomes closer to the ellipse, with $\tilde{H}=$ constant.}
\label{harmonic_well_phase20}
\end{center}
\end{figure}

\begin{figure}[htbp]
\begin{center}
\includegraphics[scale=.5]{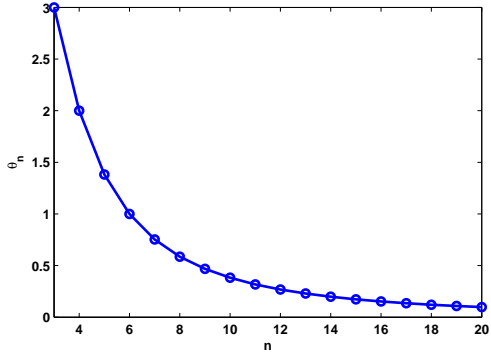}
\caption{(Color on-line) Variations of $\theta_n$ with the cycle number $n$. It is a monotonically decreasing function of $n$. As $n \to \infty$, $\theta_n \to 0$, which implies $\tau_n \to 0$.}
\label{harmonicwell01n}
\end{center}
\end{figure}

\begin{figure}[htbp]
\begin{center}
\includegraphics[scale=.5]{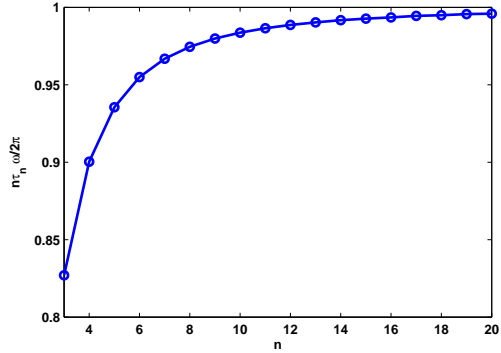}
\caption{(Color on-line) Variations of $n \tau_n$ with the cycle number $n$, showing that $n\tau_n$ is not the same as the period of the oscillator $2\pi/\omega$ in smaller $n$. As $n \to \infty$, however, $n \tau_n \to \omega/2 \pi$}.
\label{harmonicwell02n}
\end{center}
\end{figure}

\begin{figure}[htbp]
\begin{center}
\includegraphics[scale=.5]{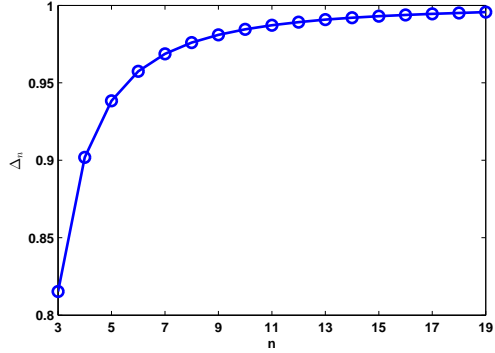}
\caption{(Color on-line) Variations of $\Delta_n$ with the cycle number $n$. Energy spacings are not even in smaller $n$. As $n \to \infty$, $\Delta_n \to 1$, which means that the calculated spacings of the energy levels are the same as the full quantum mechanical result in the limit of large $n$.}
\label{harmonicwell04n}
\end{center}
\end{figure}

In Figs. \ref{fig:01}-\ref{harmonic_well_phase20}, we use the reduced coordinates $p' \equiv p/k \tau_n q_n$, $q' \equiv q/q_n$ for each cycle $n$. We choose $(q', p') =(1, 0)$ as the initial phase points for each cycle. The phase orbit rotates clockwise on the ellipse of the modified Hamiltonian $\tilde{H}=$ constant, which is an expanded and rotated version of the original ellipse $H =$ constant. For a given cycle $n$, the discrete phase points make a closed orbit in phase space and they consequently form an $n$-polygon, as shown in Figs. \ref{fig:01}-\ref{harmonic_well_phase20}. The action integral $\oint p_ndq_n$ for the $n$-th orbit is the enclosed area of the $n$-polygon in phase space. The nonzero value of action $\oint p_ndq_n$ starts from $n=3$. A closed phase trajectory with cycle $n=3$ has $\theta_3 = 3$. The enclosed area of the triangle is given as $\oint p_ndq_n = \frac{3}{2}k \tau_3 q_3 ^2$ (Fig. \ref{fig:01}). For $n=4$, we obtain $\theta_4 = 2$ and the enclosed area of the tetragon, $\oint p_ndq_n = \frac{4}{2}k \tau_4 q_4 ^2$ (Fig. \ref{fig:02}). For larger cycle numbers, for example, $n=20$, the enclosed area of the $n$-polygon is $\oint p_ndq_n = \frac{20}{2}k \tau_{20} q_{20} ^2$ (Fig. \ref{harmonic_well_phase20}). Thus, we can deduce that the enclosed area of the $n$-polygon can be written as $\oint p_ndq_n = \frac{n}{2}k \tau_{n} q_{n} ^2$, in general. As $n$ becomes larger, the $n$-polygon becomes closer to the corresponding ellipse of the modified Hamiltonian. For the given initial condition $(q', p') =(1, 0)$, the initial total energy of the harmonic oscillator is given as $E_n = \frac{1}{2}k{q_n}^2$ for each cycle $n$ and thus we obtain a relation, $\oint p_ndq_n = n \tau_{n} E_{n}$.

Bohr \cite{bohr} proposed that an electron in a hydrogen atom can exist only in certain well-defined stable orbits, which satisfy the Bohr-Sommerfeld quantization condition \cite{sommer},
\begin{equation}
\oint p_ndq_n =nh.
\label{qcond}
\end{equation}
Note that the integral in Eq. (\ref{qcond}) is done on the $\tilde{H}=$ constant. If we impose the quantization condition on the calculated action as $\oint p_ndq_n = n \tau_{n} E_n = n h$, we finally get a relation,
\begin{equation}
E_n \tau_n = h.
\label{elevel}
\end{equation}
The product $E_n \tau_n$ is independent of the cycle number $n$ and also independent of system properties such as $k$ or $\omega$.

The calculated results for $\theta_n$ are shown in Fig. \ref{harmonicwell01n}. $\theta_n$ is a monotonically decreasing function of $n$, satisfying $0<\theta_n<4$. In fact, the parameter $\theta_n$ characterizes the quantum behaviour of a system. The larger $\theta_n$, the more the quantum behaviour. As $n \to \infty$, we have $\theta_n \to 0$ and $\tau_n\rightarrow 0$, so this corresponds to the continuous-time system. In this limit, the $n$-polygon approaches the ellipse with $\tilde{H}=$ constant. Thus, we have $n \tau_n = 2\pi/\omega$ in the limit of large $n$, \[\lim_{n \to \infty} \frac{n \omega \tau_n}{2 \pi} =1. \] This behaviour can be seen in Fig. \ref{harmonicwell02n}. $n\tau_n$ is a cycle time of the $n$-polygon, but it is not the same as the period of the oscillator $2 \pi/\omega$ for lower values of $n$. We can also rewrite Eq. (\ref{elevel}) as \[E_n =\frac{h \omega}{\sqrt{\theta_n}},\] so the spacings of the energy levels are given as $\Delta_n = 2\pi (1/\sqrt{\theta_{n+1}}- 1/\sqrt{\theta_{n}})$. The calculated results are shown in Fig. \ref{harmonicwell04n}. $\Delta_n$ rapidly increases as $n$ increases, and become less than 1 \% after $n=15$. We can see that  $\Delta_n = 1$ in the limit of large $n$. In our formulation, the result becomes the same as the full quantum mechanical result in the limit of large $n$. When $n< 6$, the energy spacings are significantly different from the full quantum mechanical results, by up to 18\%. Therefore, we can readily check the validity of our theory, if we precisely measure these differences experimentally. The experiments should be done at very low temperatures.

Other choices for the initial conditions are shown to have the same results, if we define the initial total energy $E_n$ appropriately: for example, for a different choice of the initial condition $(q', p') =(0, 1)$, $E_n \equiv \frac{1}{2}{p_n}^2$ for each cycle $n$.

Our formulation reproduces well the quantum behaviors of the particle starting from the classical mechanics. The classical phase trajectory gives us discrete energy levels without explicitly solving the Schr\"{o}dinger's equation. Since the stationary orbit starts from $n=3$, our results do not manifest any zero-point motion of the harmonic oscillator, so the zero-point motion may be a very peculiar property of the full quantum mechanics. As $n \to \infty$, $\tau_n$ goes to infinitesimally small, so the discrete system eventually turns into a continuous system. This description of the discrete particle dynamics is no longer an approximation to the continuous time dynamics and should be taken as an actual dynamics describing the quantum behaviors of the classical particle.

In summary, we obtain the discrete energy levels of the classical trajectory without explicitly solving Schr\"{o}dinger's equation. This is done by imposing the stationary condition on the phase points generated by a symplectic algorithm for a particle moving in a simple harmonic potential well. We can determine the time step $\tau_n$ and obtain the action $\oint p_ndq_n = n \tau_n E_n$ for cycle number $n\geq 3$. Additionally, by imposing the Bohr-Sommerfeld quantization condition, we obtain the discrete energy levels as $E_n \tau_n = h$. This result is based on the assumption that time is discrete. Needless to say, the varying $\tau_n$ plays a central role. A difference appears in the energy spacing $\Delta E_n \equiv E_{n+1} -E_n$ which is not evenly distributed, but as $n \to \infty$, we obtain even spacing of energy levels as $\Delta E_n = \hbar \omega$, which is quite the same as the full quantum mechanical treatment. Our theory can be checked experimentally by examining these uneven spacings of the energy levels at lower values of $n$. Our formulation introduces a new way of understanding the quantum behaviors from a different perspective and integrates classical mechanics and quantum mechanics into a single framework. We can now begin to properly understand quantum mechanics. Furthermore, we also have to identify further implications of these findings and extend them to other systems.

%%%%%%%%%%%%%%%%%%%1

\end{document}